\documentclass[aps,prc,twocolumn,showpacs,superscriptaddress,longbibliography,floatfix,10pt]{revtex4-1}
\usepackage{graphicx}
\usepackage{dcolumn}
\usepackage{bm}
\usepackage{xcolor}
\usepackage{amsfonts}
\usepackage{amssymb}
\usepackage{amsmath}
\usepackage[colorlinks,linkcolor=blue]{hyperref}

\begin{document}

\title{Shape evolution and collective dynamics of quasifission in TDHF}

\author{A.S. Umar}
\author{V.E. Oberacker}
\affiliation{Department of Physics and Astronomy, Vanderbilt University, Nashville, Tennessee 37235, USA}
\author{C. Simenel}
\affiliation{Department of Nuclear Physics, RSPE, Australian National University, Canberra, ACT 0200, Australia}
\date{\today}

\begin{abstract}
\begin{description}
\item[Background]  
At energies near the Coulomb barrier, capture reactions in heavy-ion collisions result either in fusion or in quasifission. 
The former produces a compound nucleus in statistical equilibrium, 
while the second leads to a reseparation of the fragments after partial mass equilibration without formation of a
 compound nucleus. 
Extracting the compound nucleus formation probability is crucial to predict superheavy-element formation cross-sections. 
It requires a good knowledge of the fragment angular distribution which itself depends on quantities such as  moments of inertia and excitation energies which have so far been somewhat arbitrary for the quasifission contribution.
\item[Methods]     
We investigate the evolution of the nuclear density in time-dependent Hartree-Fock
(TDHF) calculations leading to quasifission. Our main goal is to extract
ingredients of the formula used in the analysis of experimental angular
distributions.
These include the moment-of-inertia and temperature.
We study the dependence of these quantities on various initial conditions of
the reaction process.
\item[Results]     
The evolution of the  moment of inertia is clearly non-trivial and depends strongly on the characteristics of the collision. 
The temperature rises quickly when the kinetic energy is transformed into internal excitation. Then, it rises slowly during mass transfer. 
\item[Conclusions] 
Fully microscopic theories are useful to predict the complex evolution of quantities required in macroscopic models of quasifission. 

\end{description}

\end{abstract}
\pacs{21.60.-n,21.60.Jz}
\maketitle

\section{Introduction}
The creation of new elements is one of the most novel and challenging research
areas of nuclear physics~\cite{armbruster1985,hofmann1998,hofmann2000,oganessian2007}.
The search for a region of the nuclear chart that can sustain the so
called \textit{superheavy elements} (SHE) has led to intense experimental activity
resulting in the discovery and confirmation of elements with atomic numbers as large as
$Z=117$~\cite{oganessian2010,oganessian2012,hinde2014}.
The theoretically predicted \textit{island of stability}
in the SHE region of the nuclear chart is the result of new
proton and neutron shell-closures, whose location is not precisely
known~~\cite{bender1999,staszczak2013,cwiok2005}.
The experiments to discover these new elements are notoriously difficult, with
fusion evaporation residue (ER) cross-section in pico-barns.
This cross-section is commonly expressed in the product form
\begin{equation}
\sigma _{ER}=\sum_{J=0}^{J_{\max }}\sigma
_{cap}(E_{c.m.},J)P_{CN}(E*,J) W_{sur}(E*,J)
\label{eq:er}
\end{equation}
where $\sigma _{cap}(E_{c.m.},J)$ is the capture cross section at center of mass energy
E$_{c.m.}$ and spin $J$. P$_{CN}$ is the probability that the composite
system fuses into a compound nucleus (CN) rather than breaks up via quasifission. 
W$_{sur}$ is
the survival probability of the fused system against fission.
For light and medium mass systems the capture cross-section may be
considered to be the same as that for complete fusion. However, for heavy systems
leading to superheavy formation, the ER cross-section is dramatically reduced due
to the quasi-fission and fusion-fission processes~\cite{sahm1984,schmidt1991}, thus making the capture
cross-section to be essentially the sum of these two cross-sections.

The fusion process implies a transition from a dinuclear configuration 
-- accompanied by particle exchange during the dynamical process -- 
to a single-center compound-like configuration of the composite system. 
Most dynamical models~\cite{fazio2005,adamian2003,nasirov2009a,adamian2009,feng2009,zagrebaev2007,aritomo2009,zagrebaev2007c,umar2008a}
argue that, for heavy systems, a dinuclear complex is formed initially and the
barrier structure and the excitation energy of this precompound
system  determine its survival to breaking up via quasi-fission.
Furthermore, if the nucleus survives this initial state and evolves to a
compound system, it can still fission due to its excitation.

Among the three stages towards ER given in Eq.~\eqref{eq:er}, the determination of
P$_{CN}$ contains the most uncertainty which can be as
much as 1$-$2 orders of magnitude~\cite{loveland2007,yanez2013}.
Experimentally, P$_{CN}$ can be extracted from the measurement of fusion-evaporation residue cross-sections~\cite{berriman2001,khuyagbaatar2012}. 
However, these cross-sections become very small for heavy systems and the extraction of $P_{CN}$ is uncertain for such systems. 
Information on  $P_{CN}$ can also be obtained by comparing the width of fragment mass distribution with the width expected in the case of pure fusion-fission \cite{lin2012,hammerton2015}. 
This approach, however, only provides an upper limit for $P_{CN}$ 
as it assumes that only fusion-fission produces symmetric fragments \cite{lin2012}.

An alternative approach involves the analysis of the
fragment angular distribution.
For instance, assuming that the quasifission process, as fusion-fission, is statistical, 
the critical angular momentum $J_{CN}$ between fusion-fission and quasifission 
can be adjusted to reproduce the experimental angular distribution of symmetric 
fragments~\cite{back1985,tsang1983,yanez2013}. 
Although this approach has been widely used, it could have limitations due to the fact 
that quasifission in not a statistical decay but a dynamical process, 
and that fusion-fission could be asymmetric due to, e.g., late-chance fusion-fission at low excitation energies \cite{khuyagbaatar2015} as well as shell structure of pre-scission configurations \cite{mcdonnell2014}.
Correlations between mass and scattering angle of the fragments 
have also been measured extensively \cite{toke1985,rietz2011,rietz2013,wakhle2014,hammerton2015}.  
In particular, they can be used to disentangle fast quasi-fission processes (few zeptoseconds) 
to longer reaction mechanisms associated with contact times between the fragments exceeding $10-20$~zs (i.e., long-time quasifission and fusion-fission). 
However, the analysis of such fragment mass-angle distributions \cite{toke1985,rietz2011}, 
as well as statistical descriptions of fragment angular distributions \cite{back1985}, require  
external parameters such as 
 moment of inertia and temperature which are chosen somewhat
arbitrarily~\cite{back1985,yanez2013,loveland2015}.
It is therefore important to provide realistic evaluation of these parameters, in particular for the quasifission mechanism which requires a description of the complex nuclear dynamics. 

Dynamical microscopic approaches are a standard tool to extract macroscopic properties in heavy-ion collisions \cite{umar2006b,washiyama2008,wen2013,simenel2013a,simenel2013b}. 
In particular, the time-dependent Hartree-Fock theory \cite{dirac1930}, has been recognized for its realistic description of several low-energy nuclear reaction mechanisms \cite{negele1982,simenel2012}. 
It has been recently utilized for studying the dynamics of
quasifission~\cite{wakhle2014,oberacker2014,umar2015,hammerton2015}
and scission~\cite{simenel2014a,scamps2015,goddard2015}. 
The study of quasifission is showing a great promise to provide
insight based on very favorable comparisons with experimental data \cite{wakhle2014,hammerton2015}.
As a dynamical microscopic theory, TDHF provides us with the complete shape evolution of the nuclear densities 
which can be used to compute the time evolution of deformation and inertia parameters.
Using an extension to TDHF, the so-called density constraint TDHF (DC-TDHF)
approach~\cite{umar2006b},  it is also possible to compute the excitation energy in the fragments. 

In this manuscript we focus our discussion on the extraction of the time evolution of the moment of inertia 
and of the excitation energy (temperature). 
These are indeed the relevant quantities for
fragment mass-angle distributions and angular distribution analyses. 
These quantities are computed for collisions between calcium isotopes and actinides for which the TDHF approach has been shown to provide a deep insight into the quasifission reaction mechanisms \cite{wakhle2014,oberacker2014,umar2015}. 

\section{Formalism}

\subsection{TDHF and DC-TDHF approaches}

The theoretical formalism for the
microscopic description of complex many-body quantum systems
and the understanding of the nuclear interactions that result in
self-bound, composite nuclei possessing the observed properties
are the underlying challenges for studying low energy nuclear physics.
The Hartree-Fock approximation
and its time-dependent generalization, the time-dependent Hartree-Fock
theory, have provided a possible means to study the diverse phenomena
observed in low energy nuclear physics~\cite{negele1982,simenel2012}.
In general modern TDHF calculations provide a useful foundation for a
fully microscopic many-body description of large amplitude collective
motion including collective surface vibrations and giant
resonances~\cite{blocki1979,stringari1979,umar2005a,maruhn2005,nakatsukasa2005,simenel2003,reinhard2006,reinhard2007,pardi2013,pardi2014,suckling2010,stetcu2011,avez2013,scamps2014} 
nuclear reactions in the
vicinity of the Coulomb barrier, such as fusion~\cite{bonche1978,flocard1978,simenel2001,washiyama2008,umar2010a,guo2012,keser2012,simenel2013b,umar2014a,jiang2014},
deep-inelastic reactions and transfer~\cite{koonin1977,simenel2010,simenel2011,umar2008a,sekizawa2013,scamps2013,sekizawa2015},
and dynamics of (quasi)fission fragments~\cite{wakhle2014,oberacker2014,simenel2014a,umar2015,scamps2015,goddard2015}.

Despite its successes, the TDHF approach has important limitations.
In particular, it assumes that the many-body state remains a single Slater determinant at all time.
It describes the time-evolution of an independent particle system in a single mean-field corresponding to the dominant reaction channel.
As a result, it induces a classical behavior of many-body observables. 
A well known example is that TDHF does not include tunneling of the many-body wave-function and, thus, it is unable to describe sub-barrier fusion. 
It also  underestimates width of fragment mass and charge distributions in strongly dissipative collisions \cite{koonin1977,balian1984,simenel2011}. 
Thus, to obtain multiple reaction channels or widths of
one-body observables one must in principle go beyond TDHF~\cite{marston1985,bonche1985,tohyama2002a,simenel2011,lacroix2014}.

In fact, different reaction outcomes can also be obtained from mean-field calculations 
if the initial state (defined by the center of mass energy $E_{c.m.}$ 
and by the orbital angular momentum $L$ of the collision) is best described 
by a superposition of independent (quasi)particle states and that each of these states 
can be assumed to  evolve in its own mean-field. 
This is the case, for instance, if a collision partner is deformed in its intrinsic frame. 
In this case, each orientation of the deformed nucleus may encounter 
a different mean-field evolution \cite{simenel2004,umar2006d}.
Such orientation dependence of reaction mechanisms has been experimentally studied in quasifission with actinide targets
\cite{hinde1995,liu1995,hinde1996,oganessian2004a,knyazheva2007,hinde2008,nishio2008} 
and confirmed in TDHF studies \cite{simenel2012,wakhle2014,hammerton2015}.
In particular, these studies have shown that collisions of a spherical projectile 
with the tip of prolately deformed actinides lead to fast quasifission (with contact time smaller than 10~zs)
while collisions with the side of the actinide may induce longer contact times, larger mass transfer, and possible fusion. 
It is therefore common practice to investigate a subset of specific orientations 
depending on the reaction mechanism one is interested to investigate. 
For instance, it is sufficient to study ''side collisions'' in order to investigate 
the competition between quasifission and fusion in collisions with actinide nuclei \cite{oberacker2014}.

In recent years has it become numerically feasible to perform TDHF calculations on a
3D Cartesian grid without any symmetry restrictions
and with much more accurate numerical methods~\cite{umar1991a,nakatsukasa2005,umar2006c,sekizawa2013,maruhn2014,simenel2012}.
In addition, the quality of effective interactions has been substantially
improved~\cite{chabanat1998a,guichon2006,kluepfel2009,kortelainen2010}. 
In order to overcome the lack of quantum tunneling preventing direct studies of sub-barrier fusion, 
the DC-TDHF method was developed to compute
heavy-ion potentials~\cite{umar2006b} and excitation energies~\cite{umar2009a} directly from TDHF
time-evolution. For instance, this method was applied
to calculate capture cross-sections for
hot and cold fusion reactions leading to superheavy element $Z=112$~\cite{umar2010a}.

\subsection{Fragment angular distributions}

Experimental analysis of the fragment angular distributions $W(\theta)$  is commonly expressed in terms of
a two-component expression for fusion-fission and quasifission parts~\cite{huizenga1969,back1985,back1985a,keller1987},
\begin{equation}
W(\theta) = \sum_{J=0}^{J_{CN}}\mathcal{F_J^{(FF)}}(\theta) + \sum_{J=J_{CN}}^{J_{max}}\mathcal{F_J^{(QF)}}(\theta),
\label{eq:wtheta0}
\end{equation}
where
\begin{widetext}
\begin{equation}
\mathcal{F}_J^{(\alpha)}=\frac{(2J+1)^{2}exp
	[-(J+1/2)^{2}sin ^{2}\theta /4K_{0}^{2}(\alpha)]J_{0}[i(J+1/2)^{2}sin ^{2}\theta
	/4K_{0}^{2}(\alpha)]}{erf[(J+1/2)/(2K_{0}^{2}(\alpha))^{1/2}]}
\label{eq:wtheta}
\end{equation}
\end{widetext}
and $\alpha\equiv FF$ (fusion-fission) or QF (quasi-fission). 
Here, $J_{CN}$ defines the boundary between fusion-fission and quasifission, 
assuming a sharp cutoff between the angular momentum distributions of each mechanism.
The detailed definition of various mathematical functions can be found in Refs.~\cite{tsang1983,yanez2013}.

The quantum number $K$ is known to play an important role in fission \cite{vanbenbosch1973}. 
The latter is defined as the projection of the total angular momentum along the deformation axis.
In the Transition State Model (TSM)~\cite{vanbenbosch1973}, 
the characteristics of the fission fragments are determined by the $K$ distribution at scission. 
The argument $K_{0}$ entering Eq.~\ref{eq:wtheta} is the width of this distribution which is assumed to be Gaussian.
It obeys
\begin{equation}
K_{0}^{2}=T\Im _{eff}/\hbar ^{2}\;,
\label{eq:k0}
\end{equation}
where the effective moment of inertia, $\Im _{eff}$, is computed from the
moments of inertia for rotations around the axis parallel and perpendicular to the 
principal deformation axis
\begin{equation}
\frac{1}{\Im _{eff}}=\frac{1}{\Im _{\parallel }}-\frac{1}{\Im _{\perp }}\;,
\label{eq:ieff}
\end{equation}
and $T$ is the nuclear temperature at the saddle point.
The physical parameters of the fusion-fission part are relatively well known
from the liquid-drop model~\cite{sierk1986,cohen1974}.
In contrast, the quasifission process never reaches statistical equilibrium.
In principle, it has to be treated dynamically, while equation~\ref{eq:wtheta} is based on a statistical approximation.
In addition, the usual choice for the nuclear moment of inertia for
the quasifission component, $\Im _{0}/\Im _{eff}=\text{1.5}$~\cite{back1985,yanez2013,loveland2015}, is
somewhat arbitrary. Here, $\Im _{0}$ is the moment of
inertia of an equivalent spherical nucleus.

\begin{figure}[!htb]
	\includegraphics*[width=8.6cm]{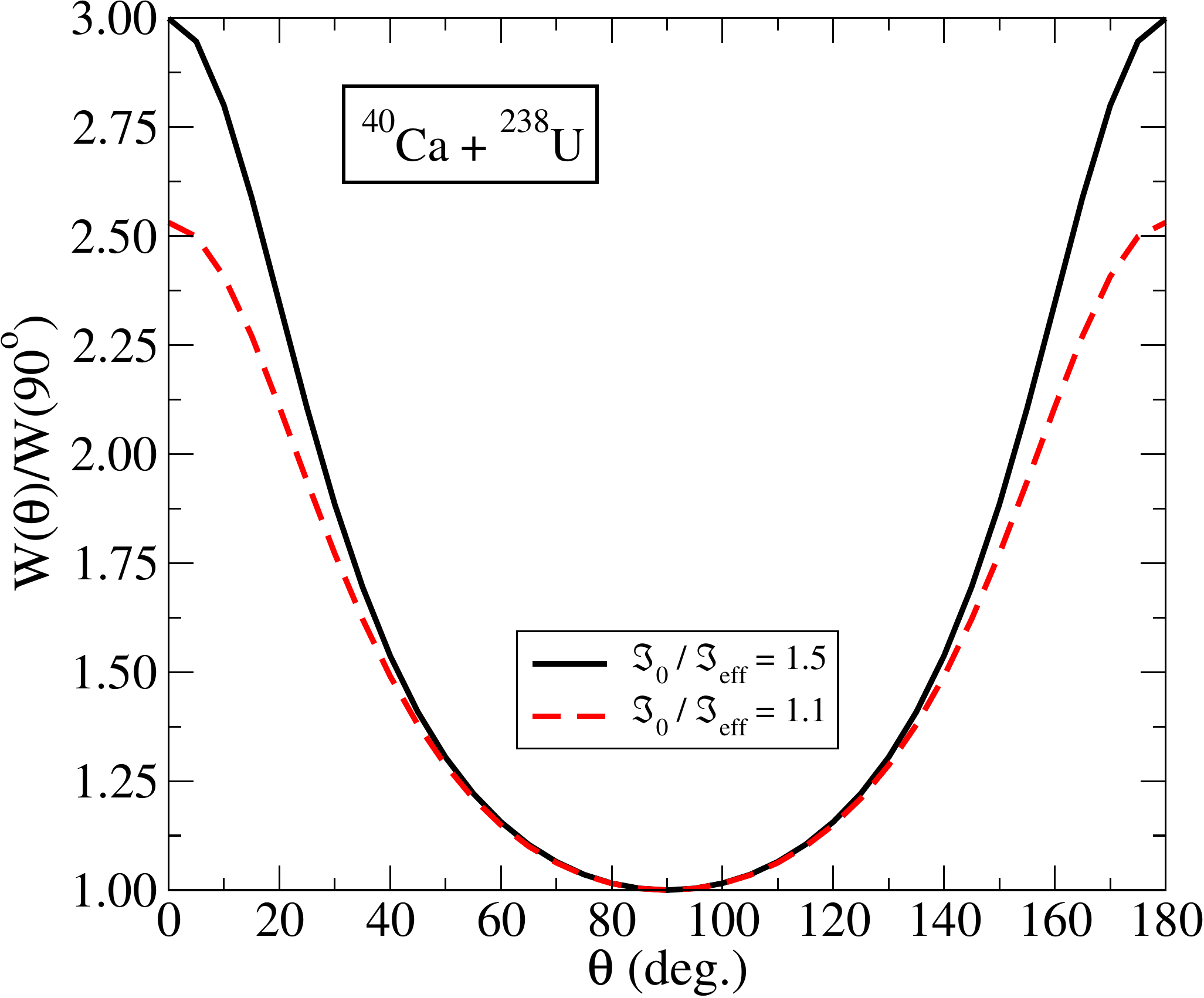}
	\caption{\protect(Color online) Effect of the moment of inertia on the angular distribution of the fragments computed with Eqs.~\ref{eq:wtheta0} and \ref{eq:wtheta}. Both curves were calculated using $J_{CN}=46$ and $T=1.2$~MeV.}
	\label{fig:W_compare}
\end{figure}
In order to illustrate the sensitivity of the fragment angular distribution with the chosen value of the moment of inertia, 
we have plotted the ratio $W(\theta)/W(90^\circ)$ in Figure~\ref{fig:W_compare} with $\Im _{0}/\Im _{eff}=\text{1.1}$ and 1.5. Here, we used the value of $J_{CN}=46$
obtained directly from TDHF calculations of quasifission for this system. Similarly,
temperature was taken to be $T=1.2$~MeV.
The deviation is mostly visible at most forward and backward angles, where it can reach up to 20$\%$.
It is interesting to note that if we assume that the red curve of Fig.~\ref{fig:W_compare} is the actual measured quantity and try to fit it by
varying the $J_{CN}$ value of the black curve we obtain a value of $J_{CN}=37$.
This would then be the error for using the wrong $J_{CN}$ value.

In the following section we outline the extraction of these ingredients directly from TDHF
time-evolution of collisions resulting in quasifission.

\section{Results}
The feasibility of using TDHF for quasifission has only been
recognized recently~\cite{simenel2012,wakhle2014,oberacker2014}.
By virtue of long contact-times for quasifission and the energy, orientation 
and impact parameter dependencies, these calculations require extremely long CPU times
and numerical accuracy~\cite{bottcher1989,umar1991a,umar2006c,maruhn2014}.
During the collision process, the nuclear densities, as described by TDHF time-evolution, undergo
complicated shape changes including vibration and rotation.
Such evolutions finally lead to two separated final
fragments identified as quasifission due to the long contact-time for the reaction
as well as the mass/charge of the fragments~\cite{oberacker2014,wakhle2014}.

\begin{figure}[!htb]
	\includegraphics*[width=8.6cm]{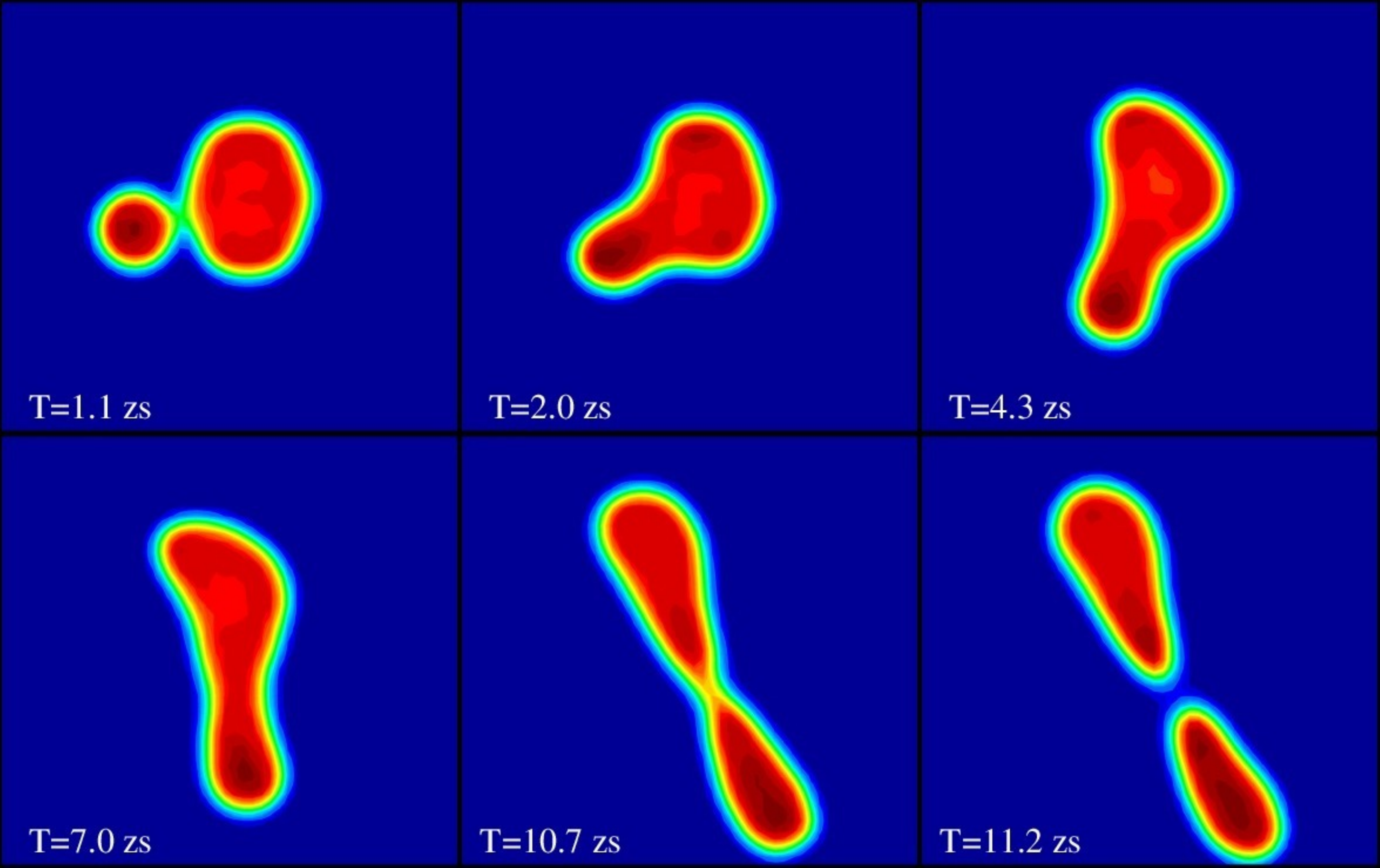}
	\caption{\protect(Color online) Quasi-fission in the reaction $^{48}$Ca+$^{249}$Bk
		at $E_{\mathrm{c.m.}}=218$~MeV with impact parameter $b=2.0$~fm.
		Shown is a contour plot of the time evolution of the mass density.
        The actual numerical box is larger than the one shown in the frames.
	}
	\label{fig:dens1}
\end{figure}

In Fig.~\ref{fig:dens1} we show a few time snapshots of the evolving mass
density for the QF reaction of the $^{48}$Ca+$^{249}$Bk system
at $E_{\mathrm{c.m.}}=218$~MeV with impact parameter $b=2.0$~fm.
The times are given in zeptoseconds (1~$zs=10^{-21}s$).
The very large elongation of the separating fragments is noteworthy.
The initial orientation of the deformed $^{249}$Bk was chosen such that
the spherical $^{48}$Ca nucleus collides with the side of the actinide
nucleus. In the following we will refer to this as the ``side''
orientation.

\subsection{Incoming and outgoing potentials}
\begin{figure}[!htb]
	\includegraphics*[width=8.6cm]{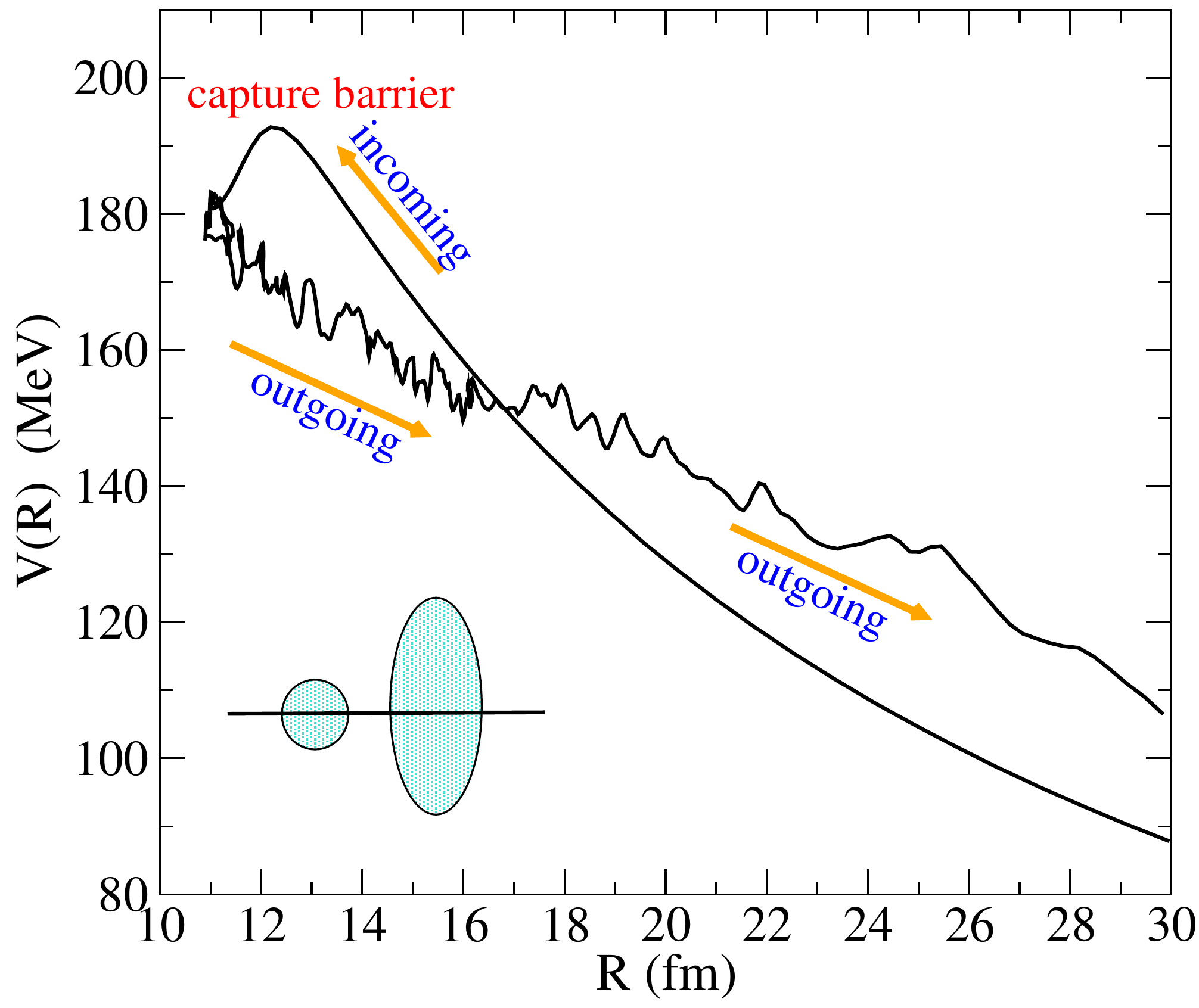}
	\caption{\protect(Color online) DC-TDHF nucleus-nucleus potential for the  $^{40}$Ca+$^{238}$U central collision
		at $E_{\mathrm{c.m.}}=211$~MeV and with the side orientation. 
		Both the potentials for the incoming and outgoing channels are shown.}
	\label{fig:pot}
\end{figure}
As quasifission is part of the capture process, 
it is therefore possible to compute the nucleus-nucleus potential 
down to the inside barrier region with the DC-TDHF technique. 
The latter is represented in Fig.~\ref{fig:pot} for the incoming channel of a $^{40}$Ca+$^{238}$U central collision.
After contact,  the two fragments encounter a significant mass transfer toward symmetry.
The outgoing potential is therefore different than the incoming one. 
In fact, the outgoing potential does not exhibit any barrier, leading to a re-separation of the fragments. 

\subsection{Moment of inertia}

The first collective observable of interest for fission and quasi-fission 
(both dynamical and statistical) studies is the moment of inertia of the system. 
The proper way to calculate the moment-of-inertia for such
time-dependent densities (particularly for non-zero impact parameters)
is to directly diagonalize the moment-of-inertia
tensor represented by a $3\times 3$ matrix with elements 
\begin{equation}
\Im_{ij}(t)/m = \int~d^3r\;\rho(\mathbf{r},t) (r^2\delta_{ij}-x_ix_j)\;,
\end{equation}
where $\rho$ is the local number-density calculated from TDHF evolution, 
$m$ is the nucleon mass, and $x_{i=1,2,3}$ denote the Cartesian coordinates. The TDHF calculations
are done in three-dimensional Cartesian geometry~\cite{umar2006c}.
Numerical diagonalization the matrix $\Im$ gives three eigenvalues.
One eigenvalue corresponds to the moment-of-inertia $\Im_{\parallel}$ for the nuclear system rotating
about the principal axis. The other two eigenvalues define the moments of inertia for 
rotations about axes perpendicular to the principal axis. 
Naturally, for triaxial density distributions, the two perpendicular components
are not exactly equal but for practical calculations they are close enough
and always larger than the parallel component.
We thus use a single average value for these moments of inertia denoted  by $\Im_{\perp}$.
\begin{figure}[!htb]
	\centering
	\includegraphics*[width=8.6cm]{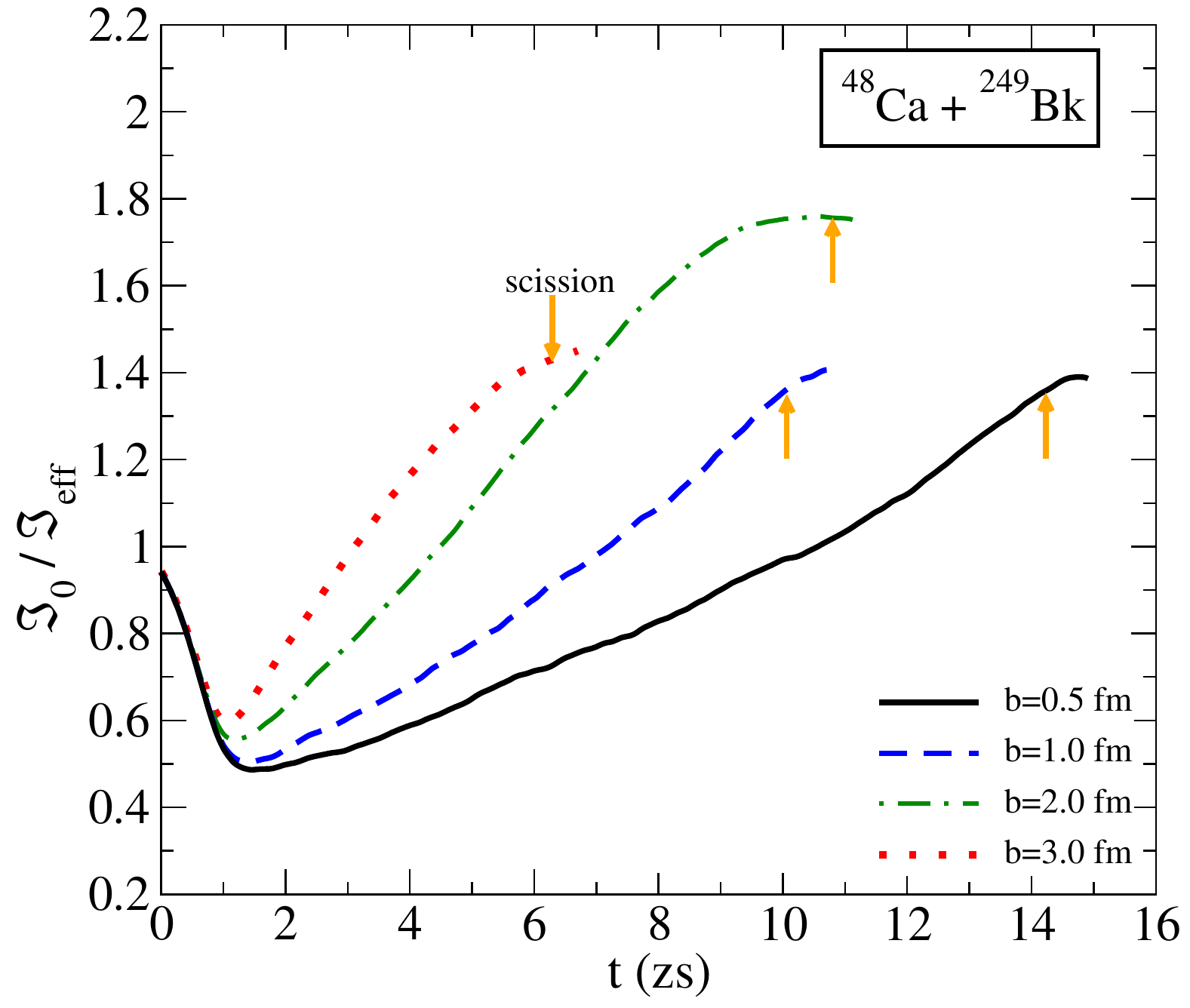}
	\caption{\protect
		TDHF results showing the time-dependence of 
		the ratio $\Im_0/\Im_{eff}$ for the $^{48}$Ca~+~$^{249}$Bk system
		at $E_\mathrm{c.m.}=218$~MeV and for impact parameters ranging from $b=0.5$ to $3$~fm.
		The arrows indicate the scission point of each trajectory. 
	}
	\label{fig:iratioB}
\end{figure}

Using the time-dependent moment-of-inertia obtained from the TDHF collision
one can calculate the so-called effective moment-of-inertia defined in
Eq.~\eqref{eq:ieff}.
It is standard to compute the effective moment of inertia relative to a spherical system 
using the mass independent quantity $\Im_0/\Im_{eff}$, 
where $\Im_0$ is the
moment-of-inertia of a spherical nucleus with the same number of nucleons~\cite{tsang1983,yanez2013}.
The expression for the moment-of-inertia for a rigid sphere is given by
$\Im_0/m=2AR_0^2/5$,
where $R_0$ can be chosen as $R_0=1.225A^{1/3}$~fm~\cite{tsang1983}.

\subsubsection{Role of impact parameter}
The moment-of-inertia ratio calculated for the $^{48}$Ca~+~$^{249}$Bk non-central collisions
at $E_\mathrm{c.m.}=218$~MeV is shown in Fig.~\ref{fig:iratioB}. 
These trajectories all lead to quasifission with a large mass transfer and orbiting before the separation of the fragments. 
This is illustrated in Fig.~\ref{fig:dens1} for the impact parameter $b=2$~fm case.
For this system using the value of $\hbar^2/m=41.471$~MeV$\cdot$fm$^2$
and total mass number $A=297$, we  get $\Im_0=191.359~\hbar^2\cdot$MeV$^{-1}$.
At the point of final touching configuration the moment-of-inertia ratios are in the range $1.4-1.8$, suggesting a
relatively strong impact parameter dependence.
Impact parameters smaller than $b=0.5$~fm lead to contact time between the fragments exceeding 35~zs. 
We can therefore consider that the nuclei have fused and would ultimately form a compound nucleus.  

\begin{figure}[!htb]
	\centering
	\includegraphics*[width=8.6cm]{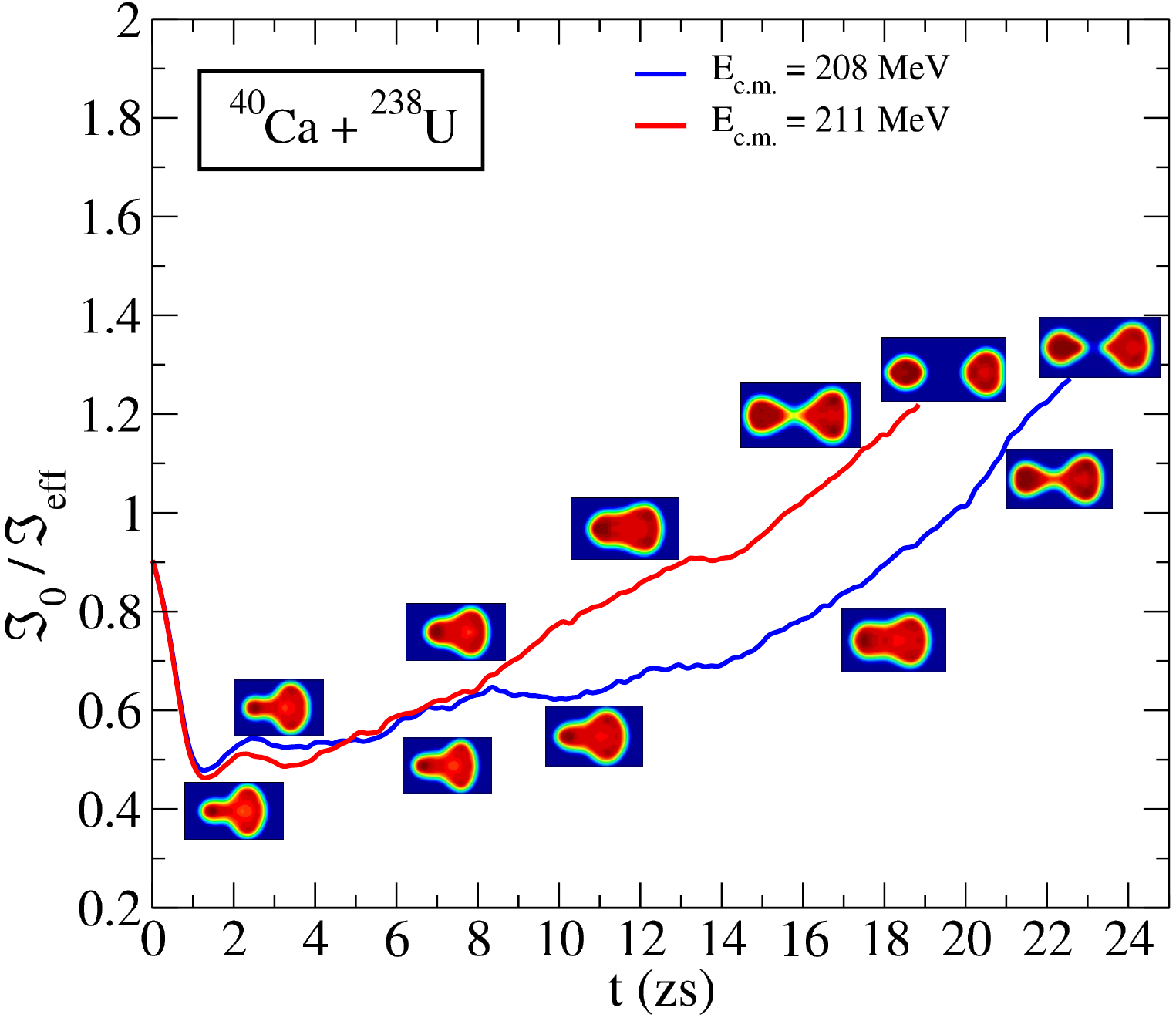}
	\caption{\protect
		TDHF results showing the time-dependence of the ratio 
		$\Im_0/\Im_{eff}$ for the $^{40}$Ca~+~$^{238}$U system
		at energies $E_\mathrm{c.m.}=208$ and $211$~MeV for zero impact parameter.
	}
	\label{fig:iratioE}
\end{figure}

\subsubsection{Role of center of mass energy}
Next we consider central collisions of the $^{40}$Ca+$^{238}$U systems.
Similar to the $^{249}$Bk case, $^{238}$U exhibits a strong prolate deformation 
and its alignment with respect to the
collision axis changes the quasifission characteristics \cite{wakhle2014}. 
The value of $\Im_0$ is 171.394~$\hbar^2\cdot$MeV$^{-1}$ for
$A=278$. 

Figure~\ref{fig:iratioE} shows the time-evolution of the moment-of-inertia ratio at
two different center of mass energies for 
central collisions with
the side orientation. 
The two calculations have an energy difference of only 3 MeV 
in order to compare reactions which lead to similar mass asymmetry of the fragments. 
It is interesting to observe that up to $\sim8$~zs, the two evolutions of the moment of inertia are almost identical. 
This is likely to be due to the fact that the translational kinetic energy is rapidly dissipated in both cases, 
leading to a system with similar compactness but different excitation energies. 
For longer times, we observe a faster re-separation of the fragments at the highest center of mass energy.
However, the final value of the ratio $\Im_0/\Im_{eff}\simeq1.2$  is approximately the same for
both energies. Note that this value is again
different from the assumed ratio of $1.5$, but in this case
 it is significantly smaller than in the  $^{48}$Ca~+~$^{249}$Bk system, 
 indicating that the $^{40}$Ca~+~$^{238}$U system leads to more compact configurations. 

\subsubsection{Role of the orientation}
\begin{figure}[!htb]
\centering
\includegraphics*[width=8.6cm]{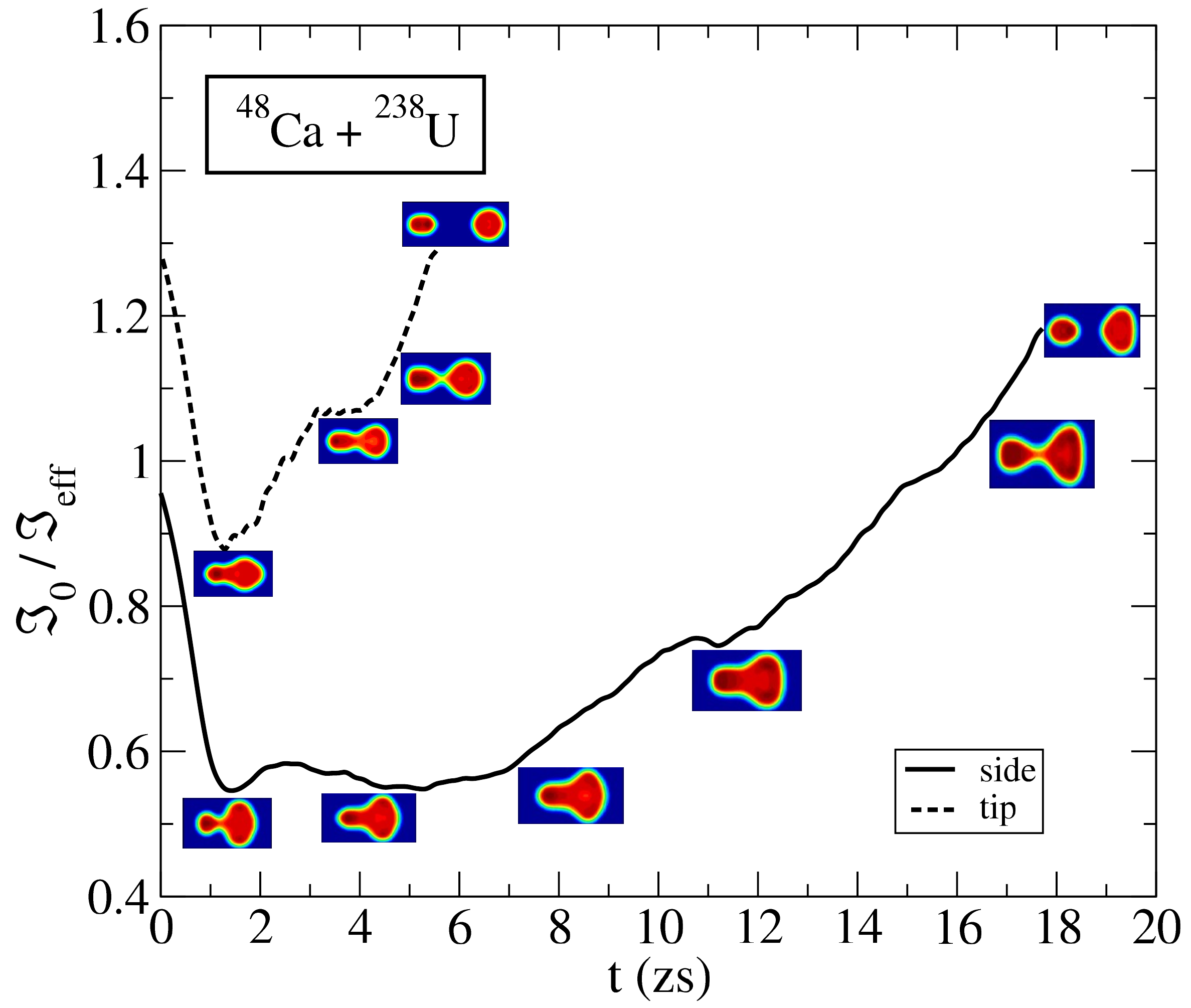}
\caption{\protect
	TDHF results showing the time-dependence of the ratio $\Im_0/\Im_{eff}$ for the $^{48}$Ca~+~$^{238}$U system
	at energy $E_\mathrm{c.m.}=203$~MeV for zero impact parameter and two
	orientations of the $^{238}$U nucleus.
}
\label{fig:iratioO}
\end{figure}

Figure~\ref{fig:iratioO} shows the dependence of the moment-of-inertia ratio on the
orientation of the $^{238}$U nucleus for the $^{48}$Ca+$^{238}$U system 
($\Im_0=179.693$~$\hbar^2\cdot$MeV$^{-1}$ for $A=286$). 
The shorter
contact time for the tip orientation (elongation axis of $^{238}$U parallel to the collision axis)
is evident from the figure.
We also observe that at the initial contact and for most of the
evolution, the ratios are significantly different, originating from the very different
initial density configurations. However, as the two reactions approach the scission point
the ratios become relatively close to one another: The neck breaks at $\Im_0/\Im_{eff}\sim1.1-1.2$.
These values are, again, significantly smaller than the assumed
value of $1.5$. 

\subsection{Collective dynamics and temperature}

In this Section we study the evolution of the TDHF density relative to the dynamical
potential energy surface (PES) using the DC-TDHF method. In the DC-TDHF method a parallel
static calculation is performed at given time intervals which constrains the instantaneous
TDHF density and finds the corresponding minimum energy state. This allows us to trace the
TDHF dynamical trajectory on the PES, albeit restricted to the shapes calculated in TDHF evolution.
At the same time these calculations provide the dynamical excitation energy, $E^{*}(t)$~\cite{umar2009a}.
The DC-TDHF calculations for heavy-systems are extremely compute extensive and
the calculation of the trajectories performed here took about two months of
computing on a $16$ processor modern workstation utilizing all processors.
\begin{figure}[!htb]
	\centering
	\includegraphics*[width=8.6cm]{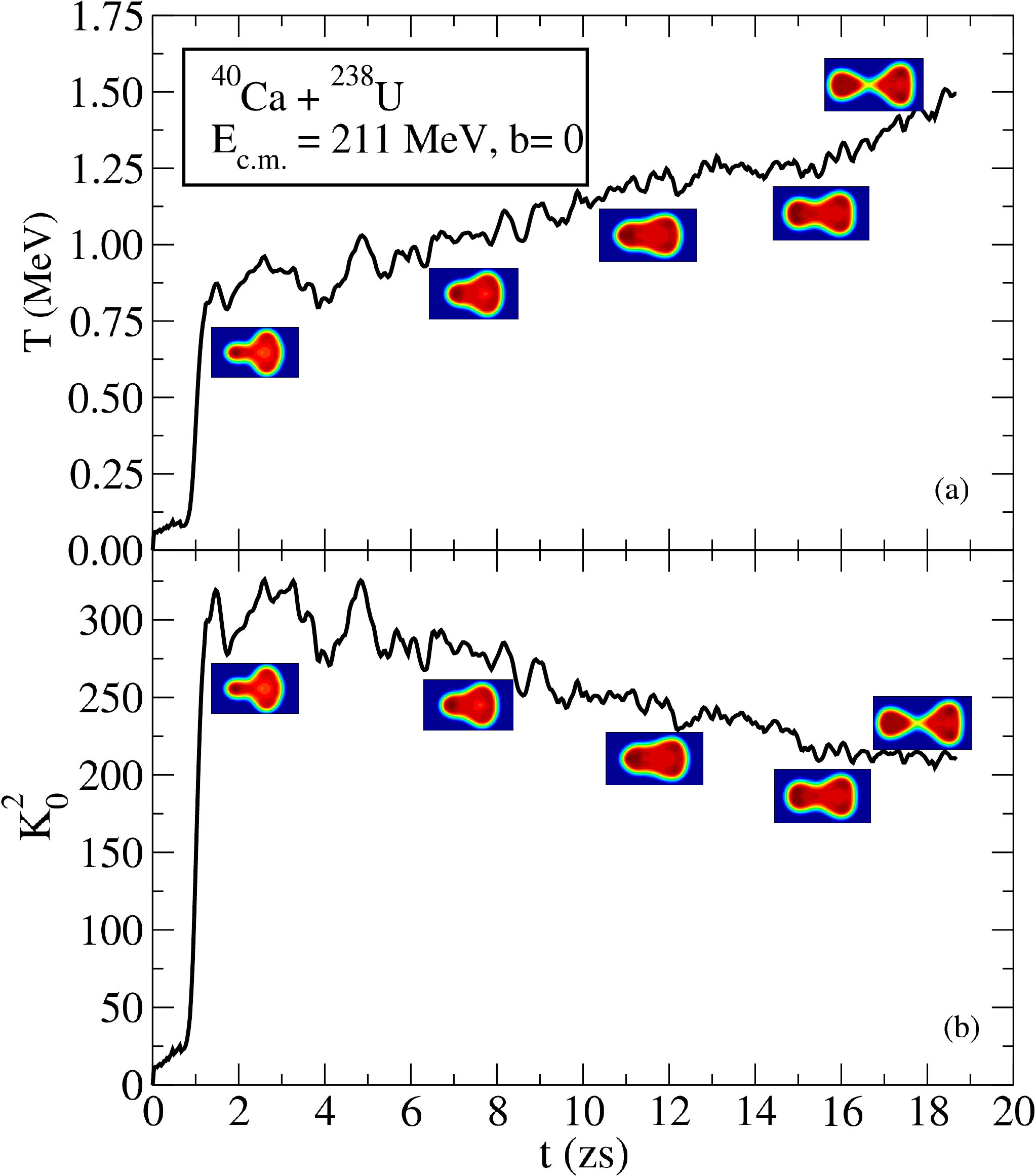}
	\caption{\protect
		TDHF results showing the time-dependence of the temperature for the $^{40}$Ca~+~$^{238}$U system
		at energy $E_\mathrm{c.m.}=211$~MeV for zero impact parameter and side
		orientation of the $^{238}$U nucleus.
	}
	\label{fig:T}
\end{figure}

The temperature used in Eq.~\eqref{eq:k0} is taken to be the temperature at the
saddle point of a fissioning system~\cite{yanez2013,loveland2015}. Since this is
adopted from modeling of fission it may not be appropriate for quasifission, which
does not really have a saddle point in the sense of ordinary fission and it is a
fully dynamical process.

We have computed the dynamical temperature of the system during the quasifission path
using the calculated excitation energy, $E^{*}(t)$, as $T(t)=\sqrt{E^{*}(t)/(A/8.5)}$ in MeV.
 Figure~\ref{fig:T}(a) shows the TDHF results 
 for the
$^{40}$Ca~+~$^{238}$U central collision at  $E_\mathrm{c.m.}=211$ MeV and side
orientation. 
We see that the temperature rapidly rises during the initial overlap phase of the collision
and reaches a value of about $1.0$~MeV. After this the temperature rises more slowly
being in the range of $1.25-1.5$~MeV along the scission path.

While the calculation of the full temperature dynamics is very computationally expensive
one can calculate the temperature if the dynamical density is stored at a chosen time.
This would then allow us to obtain all the ingredients of the parameter $K_0$ of Eq.~\eqref{eq:k0}.
One problem is that so far we do not have a convention for knowing exactly at what point
one is supposed to evaluate the ingredients of $K_0$. 
The closest configuration to a fission saddle point would be the most compact shape 
formed just after full dissipation of the translational kinetic energy. 
In this case, this would correspond to the end of the first, fast rise of $T^*(t)$, with a value of $\sim1$~MeV.

\section{Summary}
The fully microscopic TDHF theory has shown itself to be rich in
nuclear phenomena and continues to stimulate our understanding of nuclear dynamics.
We have used the TDHF theory to study evolution of the nuclear density
for reactions resulting in quasifission with a focus on the ingredients
that are used to analyze experimental angular distributions to calculate the fusion probability $P_{CN}$.
We show that a number of useful quantities can be obtained from TDHF dynamics
rather than utilizing models that may not be appropriate for quasifission.
Among these are the moment-of-inertia parallel and perpendicular to the symmetry
axis, temperature through the calculation of the dynamical excitation energy,
as well as other quantities like the rotational energy.
In addition TDHF can tell us the dependence of these variables on impact parameter,
energy, and structure.

\begin{acknowledgments}
We thank W. Loveland and D. J. Hinde for stimulating discussions.
This work has been supported by the U.S. Department of Energy under grant No.
DE-FG02-96ER40975 and DE-SC0013847 with Vanderbilt University and by the
Australian Research Council Grant No. FT120100760.
\end{acknowledgments}


\bibliography{VU_bibtex_master.bib}


\end{document}